\documentclass[prb,twocolumn,showpacs,amsmath,amssymb]{revtex4}
\usepackage{graphicx}
\usepackage{bm}

\def\br{ \bm{r} }
\def\bk{ \bm{k} }

\def\bH{ \bm{H} }
\def\bal{ \bm{\alpha} }

\begin{document}

\title{De Haas-van Alphen effect in metals without inversion center}
\author{V. P. Mineev$^{(1)}$ and K. V. Samokhin$^{(2)}$}
\affiliation{$^{(1)}$ Commissariat \'a l'Energie
Atomique, DSM/DRFMC/SPSMS 38054 Grenoble, France \\
$^{(2)}$ Department of Physics, Brock University, St.Catharines,
Ontario, Canada L2S 3A1}
\date{March 24, 2005}

\begin{abstract}
We show how the de Haas-van Alphen effect can be used to directly
measure the magnitude of spin-orbit coupling in
non-centrosymmetric metals, such as CePt$_3$Si and LaPt$_3$Si.
\end{abstract}


\maketitle

The recent discovery of superconductivity in a non-centrosymmetric
heavy-fermion compound CePt$_3$Si \cite{Bauer04} has renewed
interest, both experimental \cite{Yasuda04,Yogi04,Tate04} and
theoretical,\cite{FAKS04,SZB04,Serg04,Sam04,Min04,FAS04,Min05,Sam05}
to such materials. A peculiar property of non-centrosymmetric
metals is that the spin-orbit coupling plays an essential role in
the formation of single-electron states, namely it leads to the
splitting of the energy bands characterized by helicity (i.e. the
spin projection on the direction of momentum). This has important
consequences for superconductivity: the electrons with opposite
momenta have the same energies only if they are from the same
non-degenerate band. For electrons from different bands this is
possible only at some degeneracy lines or points in momentum
space. Therefore, a large enough band splitting prevents the
Cooper pairing of electrons from different bands.

Theoretically, the magnitude of the band splitting can be
determined from the band structure calculations. On the other
hand, one can obtain some experimental information about it from
the frequencies of de Haas-van Alphen (dHvA) oscillations of
magnetization. The first dHvA measurements in non-centrosymmetric
metals have been reported in Ref. \onlinecite{Hash04}.  While the
restoration of the Fermi surface in CePt$_3$Si is difficult due to
large values of the effective masses, the measurements on its
light-electron counterpart LaPt$_3$Si have revealed rich
information about the band structure.

Previous experimental work on the dHvA and a closely related
Shubnikov-de Haas effects in systems without inversion center
focused either on asymmetric semiconductor
heterostructures,\cite{LMFS90,ASRB94,TA02} or on bulk
semiconductors with zinc-blende structure.\cite{MR88} A common
feature of these systems is that the spin-orbit band splitting
results in two distinct frequencies of the dHvA oscillations. When
the frequencies are close, their interference produces a
characteristic beating pattern in the observed signal. This
phenomenon was first theoretically predicted in Ref.
\onlinecite{BR60} (for recent work on the subject, see e.g. Ref.
\onlinecite{WV05}). Analyzing the beating pattern allows one to
estimate the strength of the spin-orbit coupling. In this brief
article we apply these ideas to the interpretation of the dHvA
data in CePt$_3$Si and LaPt$_3$Si.

The effective single-electron Hamiltonian in a non-centrosymmetric
crystal can be written in the form
\begin{equation}
    H=\epsilon_{0}(\bk)+\bal(\bk)\bm{\sigma}
    -\mu_B\bH\bm{\sigma}, \label{e1}
\end{equation}
where $\epsilon_{0}(\bk)$ is the band energy, the spin-orbit
coupling is described by a pseudovector function $\bal(\bk)=
-\bal(-\bk)$, and $\bm{\sigma}=(\sigma_x,\sigma_y,\sigma_z)$ is
the vector composed of Pauli matrices. The last term describes the
Zeeman interaction with an external magnetic field $\bH$, with
$\mu_B$ being the Bohr magneton [using a general form of the
Zeeman energy for band electrons, $\mu_{ij}(\bk)H_i\sigma_j$,
would not add anything to the substance of our results]. The
orbital effect of the field can be included by replacing
$\bk\to\bk+(e/\hbar c)\bm{A}(\hat\br)$,\cite{LL9} where
$\hat\br=i\bm{\nabla}_{\bk}$ is the position operator in the
$\bk$-representation and $e$ is the absolute value of the electron
charge.

The momentum dependence of the pseudovector $\bal(\bk)$ is
determined by the point symmetry of the crystal. In the case of
the tetragonal group $\mathbf{C}_{4v}$, which describes the
symmetry of both CePt$_3$Si and LaPt$_3$Si, it can be written
quite generally in the form
$\bal(\bk)=\alpha_\perp[\bm{\varphi}_E(\bk)\times\hat z]+
\alpha_z\varphi_{A_2}(\bk)\hat z$, where $\bm{\varphi}_E$ and
$\varphi_{A_2}$ transform according to the irreducible
representations $E$ and $A_2$ respectively, and $\alpha_\perp$ and
$\alpha_z$ are constants.\cite{Sam04} The simplest polynomial
expression compatible with the symmetry requirements is
\begin{equation}
\label{alpha}
    \bal(\bk)=\alpha_\perp(k_y\hat x-k_x\hat y)+
    \alpha_z k_xk_yk_z(k_x^2-k_y^2)\hat z.
\end{equation}
Setting $\alpha_z=0$ here we recover the Rashba
model,\cite{Rashba60} which is used to describe the effects of the
absence of mirror symmetry in semiconductor quantum wells. In
cubic zinc-blende crystals, the momentum dependence of $\bal(\bk)$
is given by the so-called $k^3$, or the Dresselhaus,
term.\cite{Dress55,Roth68}

One cannot expect Eq. (\ref{alpha}) to fully reproduce the
spin-orbit band splitting in CePt$_3$Si and LaPt$_3$Si, which have
quite complicated, multi-sheet, Fermi surfaces. Nevertheless, this
expression already captures the most important, symmetry-related,
features of the spin-orbit coupling, including the qualitative
difference in the $\bk$-dependences of $\alpha_{x,y}(\bk)$ and
$\alpha_z(\bk)$, the presence of a band degeneracy line at
$k_x=k_y=0$, and the vanishing of $\alpha_z(\bk)$ in the
high-symmetry planes. A natural question is whether one can
determine the strengths of both the $xy$- and $z$-components of
the spin-orbit coupling using the dHvA experiments.

The eigenvalues of the Hamiltonian (\ref{e1}) are
\begin{equation}
    \epsilon_{\lambda}(\bk)=\epsilon_0(\bk)+\lambda
    |\bal(\bk)-\mu_B\bH|, \label{e3}
\end{equation}
where $\lambda=\pm $ is the band index (note that the energy bands
are split even at $H=0$, if the spin-orbit coupling is non-zero).
There are two Fermi surfaces determined by the equations
\begin{equation}
\label{e4}
    \epsilon_{\lambda}(\bk)=\epsilon_F,
\end{equation}
where $\epsilon_F$ is the Fermi energy. Although there may be
accidental degeneracies at some magnitudes and directions of the
field, in general there are no symmetry reasons for the Fermi
surfaces to intersect. Indeed, this would happen if
$\bal(\bk)=\mu_B\bH$. These three equations can have solutions at
some isolated points in the first Brillouin zone, which may or may
not be on the Fermi surface. The shape of the Fermi surfaces
(\ref{e4}) depends on the magnetic field, which can be directly
probed by dHvA experiments. In particular, while at $H=0$ we have
$\epsilon_{\lambda}(-\bk)=\epsilon_{\lambda}(\bk)$, which is a
consequence of time reversal symmetry, in the presence of magnetic
field the time-reversal symmetry is lost, and
$\epsilon_{\lambda}(-\bk)\ne\epsilon_{\lambda }(\bk)$, i.e. the
Fermi surfaces do not have inversion symmetry, in general.

To calculate the dHvA frequencies, one needs to include the
coupling of the magnetic field to the orbital motion of electrons.
In the quasi-classical approximation one can derive the
Lifshitz-Onsager quantization rules,\cite{LL9} which implicitly
determine the energy levels of the band electrons:
\begin{equation}
    S_{\lambda }(\epsilon, k_H)=\frac{2\pi eH}{\hbar c}
    \left[n+\gamma_\lambda(\Gamma)\right]. \label{e5}
\end{equation}
Here $S_\lambda$ is the area of the quasi-classical orbit,
$\Gamma$, in the $\bk$-space defined by the intersection of the
constant-energy surface $\epsilon_\lambda(\bk)=\epsilon$ with the
plane $\bk\cdot\hat{\bH}=k_H$ ($\hat{\bH}=\bH/H$), $n$ is a large
integer number, and $0\leq\gamma_\lambda(\Gamma)<1$ is a constant,
which depends on the Berry phase acquired by a band electron as it
moves along $\Gamma$.\cite{MS99,Hald04} The value of
$\gamma_\lambda(\Gamma)$ does not affect the expressions for the
dHvA frequencies discussed below.

The oscillating magnetization contains contributions from both
bands and can be approximately written as
\begin{equation}
    M_{osc}=\sum_\lambda M_\lambda\cos\left(\frac{2\pi F_\lambda}{H}
    +\phi_\lambda\right), \label{e6}
\end{equation}
where $M_\lambda$ and $\phi_\lambda$ are the amplitudes and phases
of the oscillations. The expressions for the amplitudes are given
by the standard Lifshits-Kosevich formulas.\cite{LL9} The dHvA
frequencies $F_\lambda$ are related to the extremal, with respect
to $k_H$, cross-sectional areas of the two Fermi surfaces as
follows
\begin{equation}
    F_\lambda=\frac{\hbar c}{2\pi e}S_\lambda^{ext} \label{e7}
\end{equation}
[in addition to the fundamental harmonics (\ref{e6}), the observed
dHvA signal also contains higher harmonics with frequencies given
by multiple integers of $F_\lambda$].

If the external field is weak compared to the spin-orbit band
splitting, i.e. $\mu_BH\ll|\bal(\bk)|$, the band energies
(\ref{e3}) can be represented as a Taylor expansion
\begin{eqnarray}
\label{taylor}
    \epsilon_{\lambda}(\bk)=\epsilon_{0}(\bk)+\lambda|\bal(\bk)|
    -\lambda\mu_B(\hat\bal\bH)\nonumber\\
    +\frac{\lambda\mu_B^2}{2|\bal(\bk)|}[H^2-(\hat\bal\bH)^2]+\ldots,
\end{eqnarray}
where $\hat\bal(\bk)=\bal(\bk)/|\bal(\bk)|$. Similarly, the
extremal cross-section areas can be written in the form
\begin{equation}
    S_\lambda^{ext}(\bH)=S_\lambda^{ext}(0)+A_\lambda(\hat{\bH})H+
    B_\lambda(\hat{\bH})H^{2}+\ldots
\label{e8}
\end{equation}
The second, linear in $H$, term on the right-hand side produces
the phase shifts in dHvA signal (\ref{e6}). This effect is similar
to the usual phase shift due to a paramagnetic splitting of Fermi
surfaces in centrosymmetric metals. For some directions of the
field, the linear term can be absent, see an example below. The
third term and all the subsequent terms produce the magnetic field
dependence of the dHvA frequencies. This is a specific feature of
the dHvA oscillations in crystals without inversion symmetry,
which can be observable if the Zeeman energy is not too small in
comparison with spin-orbit coupling. A non-linear field dependence
of the dHvA frequencies has been observed in asymmetric quantum
wells.\cite{LMFS90}

To illustrate the above statements, let us look at a simple
example of a three-dimensional elliptic Fermi surface with
$\epsilon_0(\bk)=\hbar^2k_\perp^2/2m_\perp+\hbar^2k_z^2/2m_z-\epsilon_F$,
where $\bk_\perp=(k_x,k_y)$, and $m_\perp,m_z$ are the effective
masses. The Fermi momentum $k_F$ is introduced via
$\epsilon_F=\hbar^2k_F^2/2m_\perp$. We consider only
$\bH\parallel\hat z$ to make connection with the experimental
results of Ref. \onlinecite{Hash04}, where two main dHvA branches,
named $\alpha$ and $\beta$, were detected for this field
orientation. One can show that the linear in $H$ terms in the
expansions (\ref{taylor}) and (\ref{e8}) vanish. The maximum
cross-sections of the Fermi surfaces correspond to $k_z=0$, then
$\varphi_{A_2}=0$ and we obtain the extremal cross-section area
which depends only on the transverse spin-orbit coupling:
\begin{equation}
    S_\lambda^{ext}(\bH)=\pi k_F^2\left[1-\lambda\frac{|\alpha_\perp|k_F}{\epsilon_F}
    \left(1+\frac{\mu_B^2H^2}{2\alpha_\perp^2k_F^2}\right)\right].
\label{cross-sec}
\end{equation}
In obtaining this result we used the expression (\ref{alpha}) for
$\bal$ and assumed that the Zeeman energy is small compared to the
spin-orbit band splitting, which in turn is much smaller than the
Fermi energy: $\mu_BH\ll|\alpha_\perp| k_F\ll\epsilon_F$.
Although, for a more complicated Fermi surface, there might be
additional extremal cross-sections at nonzero
$k_z$,\cite{BZ-faces} the linear in $H$ term in Eq. (\ref{e8}) is
still absent due to the symmetry properties of $\varphi_{A_2}$.

To estimate the magnitude of the effects under consideration, we
use the expressions (\ref{e7}) and (\ref{cross-sec}) to calculate
the difference of the dHvA frequencies:
\begin{equation}
    F_--F_+=\frac{2c}{\hbar e}|\alpha_\perp|k_Fm_\perp
    \left(1+\frac{\mu_B^2H^2}{2\alpha_\perp^2k_F^2}\right). \label{e9}
\end{equation}
The experimental measurement of the splitting of the frequencies
allows one to determine the strength of the spin-orbit coupling.
Using as an example the frequencies of the $\alpha$ and $\beta$
branches from Ref. \onlinecite{Hash04} $F_{\alpha}=1.10\times
10^8$Oe and $F_{\beta}=8.41\times 10^7$Oe, and $m_\perp\simeq
1.5m$, we obtain for the spin-orbit splitting of the Fermi
surfaces: $|\alpha_\perp|k_F\simeq 10^3$K. While the results of
the band structure calculations for LaPt$_3$Si reported in Ref.
\onlinecite{Hash04} do not contain explicit values of the band
splitting $\Delta E_{so}$, for CePt$_3$Si one has $\Delta
E_{so}\simeq 50-200$meV.\cite{SZB04} As for the magnitude of the
magnetic field dependence of the frequency splitting, in the range
of fields used in Ref. \onlinecite{Hash04} (up to $17$T), we have
$\mu_BH/|\alpha_\perp|k_F\sim 10^{-2}$.

We would like to note that the expansions (\ref{taylor}) and
therefore (\ref{e8}) fail if $\bm{\alpha}(\bk)=0$. According to
Eqs. (\ref{alpha}), this happens if the extremal orbit passes
through the poles of the Fermi surface, where the bands are
degenerate. In this case, the so-called ``magnetic breakdown''
occurs, in which the electrons can tunnel from one band to another
near the degeneracy points. Instead of Eq. (\ref{e6}), the dHvA
signal then contains additional fundamental harmonics
corresponding to the quasi-classical orbits switching between
different bands.\cite{MB} It is not clear if this phenomenon
occurs in LaPt$_3$Si and CePt$_3$Si.

In conclusion, we have discussed how the absence of inversion
symmetry in the crystal lattice of a metal manifests itself in the
dHvA experiments. The splitting of the dHvA frequencies is a
direct measure of the parameters of the effective spin-orbit
Hamiltonian. In particular, according to Eq. (\ref{e9}), it allows
one to estimate the magnitude of the ``transverse'' component of
the spin-orbit coupling (in contrast, there seems to be no simple
way to determine the $z$-axis component using the dHvA data).
Also, the interplay of the Zeeman and the spin-orbit interactions
results in a deformation of the Fermi surface, which is
responsible for a non-linear field dependence of the dHvA
frequencies, the effect absent in centrosymmetric metals.

\acknowledgements

The authors are grateful to E. I. Rashba for valuable comments and
interest to this work. One of us (V.M.) is indebted to Dr. N.
Tateiwa for drawing his attention to the recent dHvA experiments
in non-centrosymmetric metals. The support from the Natural
Sciences and Engineering Research Council of Canada (K.S.) is
gratefully acknowledged.


\begin{thebibliography}{90}
\bibitem{BZ-faces}
One can show that, in addition to the plane $k_z=0$,
$\alpha_z(\bk)$ also vanishes on the horizontal faces of the first
Brillouin zone, i.e. $k_z=\pm\pi/c$.
\end{thebibliography}
\end{document}